\documentclass[fleqn,10pt]{wlscirep}
\usepackage[utf8]{inputenc}
\usepackage[T1]{fontenc}
\usepackage{bm}

\usepackage{pdfpages}

\DeclareMathOperator*{\argmax}{arg\,max}

\title{A framework for subsurface monitoring by integrating reservoir simulation with time-lapse seismic surveys}

\author[1,*]{Johno van IJsseldijk}
\author[1]{Hadi Hajibeygi}
\author[1]{Kees Wapenaar}
\affil[1]{Delft University of Technology, Department of Geoscience and Engineering, Delft, The Netherlands}

\affil[*]{corresponding author: J.E.vanIJsseldijk@tudelft.nl}


\begin{abstract}
Reservoir simulations for subsurface processes play an important role in successful deployment of geoscience applications such as geothermal energy extraction and geo-storage of fluids. These simulations provide time-lapse dynamics of the coupled poromechanical processes within the reservoir and its over-, under-, and side-burden environments. For more reliable operations, it is crucial to connect these reservoir simulation results with the seismic surveys (i.e., observation data). However, despite being crucial, such integration is challenging due to the fact that the reservoir dynamics alters the seismic parameters. In this work, a coupled reservoir simulation and time-lapse seismic methodology is developed for multiphase flow operations in subsurface reservoirs. 
To this end, a poromechanical simulator is designed for multiphase flow and connected to a forward seismic modeller. This simulator is then used to assess a novel methodology of seismic monitoring by isolating the reservoir signal from the entire reflection response. This methodology is shown to be able to track the development of the fluid front over time, even in the presence of a highly reflective overburden with strong time-lapse variations. These results suggest that the proposed methodology can contribute to a better understanding of fluid flow in the subsurface. Ultimately, this will lead to improved monitoring of reservoirs for underground energy storage or production. 
\end{abstract}
\begin{document}

\flushbottom
\maketitle
%
%
\thispagestyle{empty}


\section*{Introduction}
Understanding fluid flow in subsurface reservoirs is crucial to predict underground processes related to the energy transition, such as geothermal energy \citep{Barbier2002}, temporary storage of green gasses like hydrogen \citep{Kumar2022}, and long-term storage of greenhouse gasses like CO$_2$ \citep{Wang2022}. Reservoir simulations allow us to accurately predict fluid flow inside a reservoir, based on a combination of geological, geophysical and borehole data \citep{Peaceman2000,Berkowitz2002}. Geophysical methods, such as seismic monitoring, are able to observe time-lapse changes of dynamic properties like pressure and fluid saturation, everywhere in a three-dimensional subsurface. Seismic monitoring relies on the fact that changes in the reservoir will translate into changes in the seismic reflection response. The fluid flow inside the reservoir can then be imaged by comparing a baseline seismic survey with a monitor survey, recorded over the same location at a later point in time \citep{Tura1998,Lumley2001,Landro2001,Johnston2013}. \\
Feasibility studies aim to assess the seismic detectability of fluid movement inside a hydrocarbon reservoir \citep{Lumley1994} or migration of injected CO$_2$ for CCS projects \citep{Pevzner2011,Macquet2019}. These types of studies rely on reservoir simulations to predict the movement of the fluids in the reservoir. Although this methodology provides accurate estimates of the time-lapse changes inside the reservoir, it does not predict geomechanical changes in the overburden. However, these changes can have large effects on the repeatability of time-lapse experiments, as overburden time-shifts might be mistaken for changes inside the reservoir \citep{Calvert2005}. Generally, an independent geomechanical model is used to compute the time-lapse changes in the layers above the reservoir \citep{Hatchell2005b,Macbeth2021}. Recently, multiphase poromechanical models were introduced as an all-in-one solution to link fluid flow, transport and deformation in the subsurface \citep{white2019}. Traditionally, these models are used to predict induced seismicity due to fluid injection in the subsurface \citep{Birendra2014,Castineira2016,Han2021}. Additionally, poromechanical simulations can, in theory, also be used to model both time-lapse changes inside the reservoir and overburden at once for seismic monitoring applications. \\
In addition to time-lapse overburden effects, static overburden effects can also obstruct the reservoir signal in the baseline and monitor seismic surveys, due to the presence of highly reflective layers in the overburden. Both the static and dynamic overburden effects can be accounted for by isolating the reservoir response \citep{WapenaarIJsseldijk2020,vanIJsseldijk2023}. This isolation is based on the 3D Marchenko equations that describe all orders of multiple scattering inside the medium \citep{broggini2012focusing,wapenaar2014marchenko,slob2014seismic}. After this Marchenko-based isolation is applied to the seismic data, the reflections related to the reservoir are clearly visible in the seismic response. Next, the primary reflection from the top of the reservoir is used as a reference event that contains all the delays of the overburden. This reference event is then combined with events originating from the reservoir's base to retrieve time-lapse differences that are solely dependent on the changes inside the reservoir \citep{vanIJsseldijk2023}. \\
In this work, a poromechanical simulator is proposed to model time-lapse changes in density and compressional velocity due to fluid injection in a subsurface reservoir. Since multiphase fluid flow as well as geomechanics are included in the formulation, the changes in the overburden and reservoir are modelled all at once. Next, the velocities and densities are computed at a number of time-steps during the simulation, which are used to model the seismic response for the seismic baseline and different monitor studies. Finally, time-lapse changes are retrieved. These changes are then independently assessed both before and after isolation of the reservoir response from the total seismic response (i.e. the response of the overburden, reservoir and underburden). In the next section, we will first discuss the governing equations behind the poromechanical model, the connection with seismic parameters and the retrieval of time-lapse traveltime differences. Secondly, the methodology is tested on a simple as well as a complex model. To conclude, we discuss the results and possible future improvements and extensions to the method.

\section*{Methods}

This section discusses the background on how time-lapse changes can be extracted from a modelled reservoir. The constitutive equations related to poromechanics are first reviewed, then these equations are related to seismic properties, which can be used to model the seismic response at different times in the simulation. These responses are compared to one another to find seismic time-lapse traveltime differences between the different surveys.

\begin{figure}[tb!]
    \centering
    \includegraphics[width=.7\textwidth]{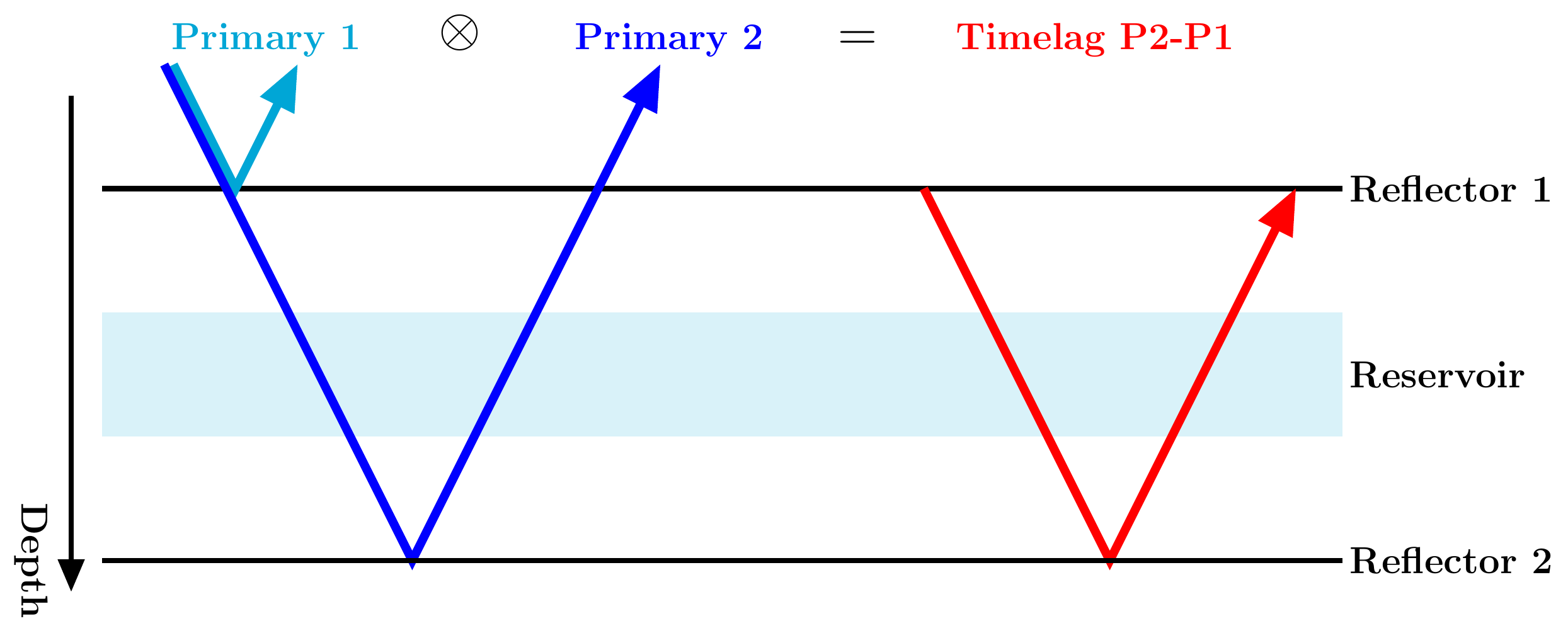}
    \caption{Principle of seismic interferometry: a reservoir is situated between two reflectors that reflect primaries 1 and 2. Cross-correlating these primaries cancels their common path, hence only the timelag between the two reflections remains (i.e. the traveltime through the reservoir).}
    \label{fig:Interferometry}
\end{figure}

\subsection*{Multiphase poromechanics}

The geomechanical changes in an isotropic subsurface are represented by the conservation of momentum  (\autoref{eqn:momentum1} \citep{Wang2001}), and conservation of mass describes flow of immiscible fluids through a reservoir (\autoref{eqn:alpha1} and \ref{eqn:beta1} \citep{Aziz1979}). This gives the following system of equations:
\begin{align}
& \nabla \cdot \bm{\sigma} = \mathbf{f}, \label{eqn:momentum1} \\
& \frac{\partial m_\alpha}{\partial t}  + \nabla \cdot (\rho_\alpha \mathbf{v}_\alpha) = \rho_\alpha q_\alpha, \label{eqn:alpha1} \\
& \frac{\partial m_\beta}{\partial t}  + \nabla \cdot (\rho_\beta \mathbf{v}_\beta) = \rho_\beta q_\beta. \label{eqn:beta1} 
\end{align}
Here, $m$, $\rho$, $\mathbf{v}$ and $q$, are the fluid mass per unit volume, density, velocity and sinks/sources, respectively. The subscripts $\alpha$ and $\beta$ denote two different fluid phases. Furthermore, $t$ denotes time, $\bm{\sigma}$ is the stress tensor, and $\mathbf{f}$ stands for the body forces acting on the system. Next, the stress tensor in \autoref{eqn:momentum1} is connected to the changes in fluid pressure and displacement according to Biot's theory of poroelasticity \citep{Biot1941}. Moreover, the mass per unit volume of each phase is related to it's saturation, density and the porosity (i.e. $m = \phi S \rho$). Finally, Darcy's law is used to write the fluid velocity in terms of phase mobility and pore pressure. After applying all these conditions, the system of equations \ref{eqn:momentum1}-\ref{eqn:beta1} now reads \citep{white2019}:
\begin{align}
& \nabla \cdot \left( \mathbb{C}_{dr} :  \nabla^s \textbf{u} - bp_{f}\textbf{I} \right) =  \mathbf{f}, \label{eqn:momentum2} \\
& \frac{\partial \phi S_\alpha \rho_\alpha}{\partial t}  - \nabla \cdot (\bm{\lambda}_\alpha \rho_\alpha \nabla p_f) = \rho_\alpha q_\alpha, \label{eqn:alpha2} \\
& \frac{\partial \phi S_\beta \rho_\beta}{\partial t}  - \nabla \cdot (\bm{\lambda}_\beta\rho_\beta \nabla p_f) = \rho_\beta q_\beta. \label{eqn:beta2}
\end{align}
In \autoref{eqn:momentum2} $b$ denotes Biot's coefficient, $\nabla^s \textbf{u} = 0.5 (\nabla \textbf{u} + \nabla \textbf{u}^T)$ is the symmetric gradient operator operating on displacement \textbf{u}, \textbf{I} is a unit tensor and $\mathbb{C}_\text{dr}$ the rank-4 drained elasticity tensor, which for isotropic linear elastic material reads:
\begin{equation}
    \mathbb{C}_{\text{dr},ijkl} = \lambda_\text{dr} \delta_{ij}\delta_{jk} + G_\text{dr} ( \delta_{ik}\delta_{jl} + \delta_{il}\delta_{jk}).
\end{equation}
Here, $\delta_{ij}$ is Kronecker's delta, $\lambda$ Lam\'e's first parameter, and $G$ Lam\'e's second parameter or the shear modulus. The subscript dr denotes that the elastic moduli are drained. $p_f$ in equations \ref{eqn:momentum2}-\ref{eqn:beta2} represents the pore pressure, which in the absence of capillary forces is the same for each phase. In \autoref{eqn:alpha2} and \ref{eqn:beta2} $S$ indicates the saturation, for two-phase flow $S_\beta = 1 - S_\alpha$. Moreover, $\bm{\lambda}$ depicts the phase mobility, which is equal to the rock permeability times the relative permeability over the viscosity ($\mathbf{K} k_r/\mu$). Lastly, $\phi$ is the porosity, which differss from a reference $\phi_0$ due to the fluid pressure and volumetric strain $\epsilon_v = \text{trace}(\nabla^s \textbf{u})$ as \citep{Coussy2004}:
\begin{equation}
    \Delta \phi = b \Delta \epsilon_v + \frac{(b-\phi_0)(1-b)}{K_\text{dr}} \Delta p_f,
\end{equation}
with drained bulk modulus $K_\text{dr}=\lambda_\text{dr} + (2/3) G_\text{dr}$.

\subsection*{Seismic parameters via fluid substitution}
After the dynamic fluid and geomechanic quantities have been computed by the poromechanical simulation, they have to be converted into seismic parameters, namely density and compressional wave velocity. Note that, in this study, only compressional waves are considered for the forward seismic modelling, even though retrieving the shear wave velocity is trivial once all elastic parameters are calculated. This is due to current limitations of the Marchenko-based isolation of the reservoir response, as the Marchenko equations are not straightforwardly applied to elastic theory \citep{Reinicke2020}, but extensions are under investigation \citep{daCosta2014,Reinicke2019}. The saturated density can be calculated using the fluid saturation and density as well as the porosity and rock density:
\begin{equation}
    \rho_\text{sat} = (1-\phi)\rho_\text{rock} + \phi(\rho_\alpha S_\alpha + \rho_\beta S_\beta).
\end{equation}
Next, the compressional wave velocity $c_p$ is computed using the elastic moduli $K$ and $G$ as well as density $\rho$: 
\begin{equation}
c_p = \sqrt{\frac{K_\text{sat}+\frac{4}{3}G_\text{sat}}{\rho_\text{sat}}},
\end{equation}
where the subscript sat denotes a saturated medium. Gassmann's equation describes how the saturated bulk and shear moduli can be found \citep{Gassmann1951,Biot1956}:
\begin{align}
& K_\text{sat} = K_\text{dr} + \frac{(1 - K_\text{dr}/K_0)^2}{\phi / K_\text{fl} + (1-\phi)/K_0 - K_\text{dr}/K_0^2}, \label{eqn:gassmann1}\\
& G_\text{sat} = G_\text{dr}. \label{eqn:gassmann2}
\end{align}
Again, $K_\text{dr}$ and $G_\text{dr}$ are the drained bulk and shear modulus, respectively \citep{Mavko2009}. $K_0$ is the bulk modulus of the minerals of the rock that can be experimentally determined \citep[e.g.][]{Manika2002,Angel2009}. $K_\text{fl}$ is effective bulk modulus of the fluid that can, for example, be calculated using the Reuss method for uniform saturation \citep{Reuss1929}:
\begin{equation}
    K_\text{fl} = \frac{1}{S_\alpha\kappa_\alpha + S_\beta\kappa_\beta},
\end{equation}
where $\kappa$ is the compressibility of the fluid, which is equal to the inverse of the bulk modulus ($\kappa = 1 / K$), and can be derived from the pressure, volume and temperature of the fluid \citep[e.g.][]{Rackett1970}. Note that \autoref{eqn:gassmann1} and \ref{eqn:gassmann2} are only valid at low frequencies ($<100$ Hz), which makes them ideal for field-scale experiments such as in this study \citep{Mavko2009}.

\subsection*{Extracting time-lapse traveltime differences}
\newcommand{\xs}[1]{\mathbf{x}_{#1}}
First, the subsurface is divided in three units, overburden $a$, target zone $b$ and underburden $c$. The reservoir is located in the target zone. Next, the velocity and density explained in the previous section are used to model the seismic reflection response $R_{abc}(\xs{R},\xs{S},t)$ of the entire subsurface, specified by the subscript $abc$. The source and receiver coordinates at the surface are indicated with $\xs{S}$ and $\xs{R}$, respectively. Additionally, $t$ denotes the seismic recording time, which is different from the flow simulation time in \autoref{eqn:alpha2} and \ref{eqn:beta2}. The seismic recording time is typically in the order of seconds, whereas the flow simulation time is in the order of hours to days. Reflections in the overburden can interfere with the signal from the reservoir, which can prevent the accurate retrieval of time-lapse differences. Therefore, the reservoir response has to be isolated from the full response. This isolated response $R_b(\xs{R},\xs{S},t)$ is free from over- and underburden reflections, which allows accurate retrieval of the time-lapse differences inside the reservoir \citep{vanIJsseldijk2023}. Details of this isolation are discussed in the supplementary material accompanying this paper. \\
Correlations are a popular method to extract time-lapse traveltime differences from seismic data \citep{Snieder2002,Macbeth2020}. The traveltime differences $\Delta t$ can be found by cross-correlating the same signal between a baseline and monitor survey and taking the argument of the maximum of the correlation:
\begin{equation}
\label{eqn:timediffs}
\Delta t (\mathbf{x}_0) = \argmax_\tau \left( \int_0^\infty C(\mathbf{x}_0,t+\tau) \bar{C}(\mathbf{x}_0,t) d t \right).
\end{equation}
In \autoref{eqn:timediffs}, $\xs{0}$ represents the zero-offset coordinate, where $\xs{S}=\xs{R}$. Moreover, $C$ denotes the event to be correlated of the baseline survey, and $\bar{C}$ the same event in the monitor survey. If this event is simply a primary originating from a reflector below the reservoir, all the time-delays of the overburden will be present in the calculated traveltime differences. Instead a control reflection from above the reservoir can be used to first compute the timelag inside the reservoir \citep{vanIJsseldijk2023}. In \autoref{fig:Interferometry} this idea is systematically depicted; primary 1 does not travel through the reservoir, while primary 2 does. By cross-correlating these two primaries the timelag in the reservoir is computed, while all time-delays from the overburden are cancelled out. This is akin to the idea of seismic interferometry \citep{Wapenaar2010}. \autoref{eqn:timelags} describes how correlation of primaries 1 ($P1$) and 2 ($P2$) returns the timelag ($C_{P1P2}$) between these two events: 
\begin{equation}
\label{eqn:timelags}
C_{P1P2}(\mathbf{x}_0,\tau) =
\int_0^\infty P1(\mathbf{x}_0,t+\tau) P2(\mathbf{x}_0,t) d t.
\end{equation}
If the timelag in \autoref{eqn:timelags} is computed for both the baseline and monitor survey, the retrieved correlations can be inserted into \autoref{eqn:timediffs} in order to acquire time-lapse traveltime differences that only encompass the changes in the reservoir layer. A similar procedure can be applied to multiples, selected from the isolated target response $R_b$. Since multiples have traveled through the reservoir layer multiple times, they are more sensitive to time-lapse changes in the reservoir \citep{vanIJsseldijk2023}. In the following, only primaries are considered.

\begin{figure}[tb!]
    \centering
    \includegraphics[width=.45\textwidth]{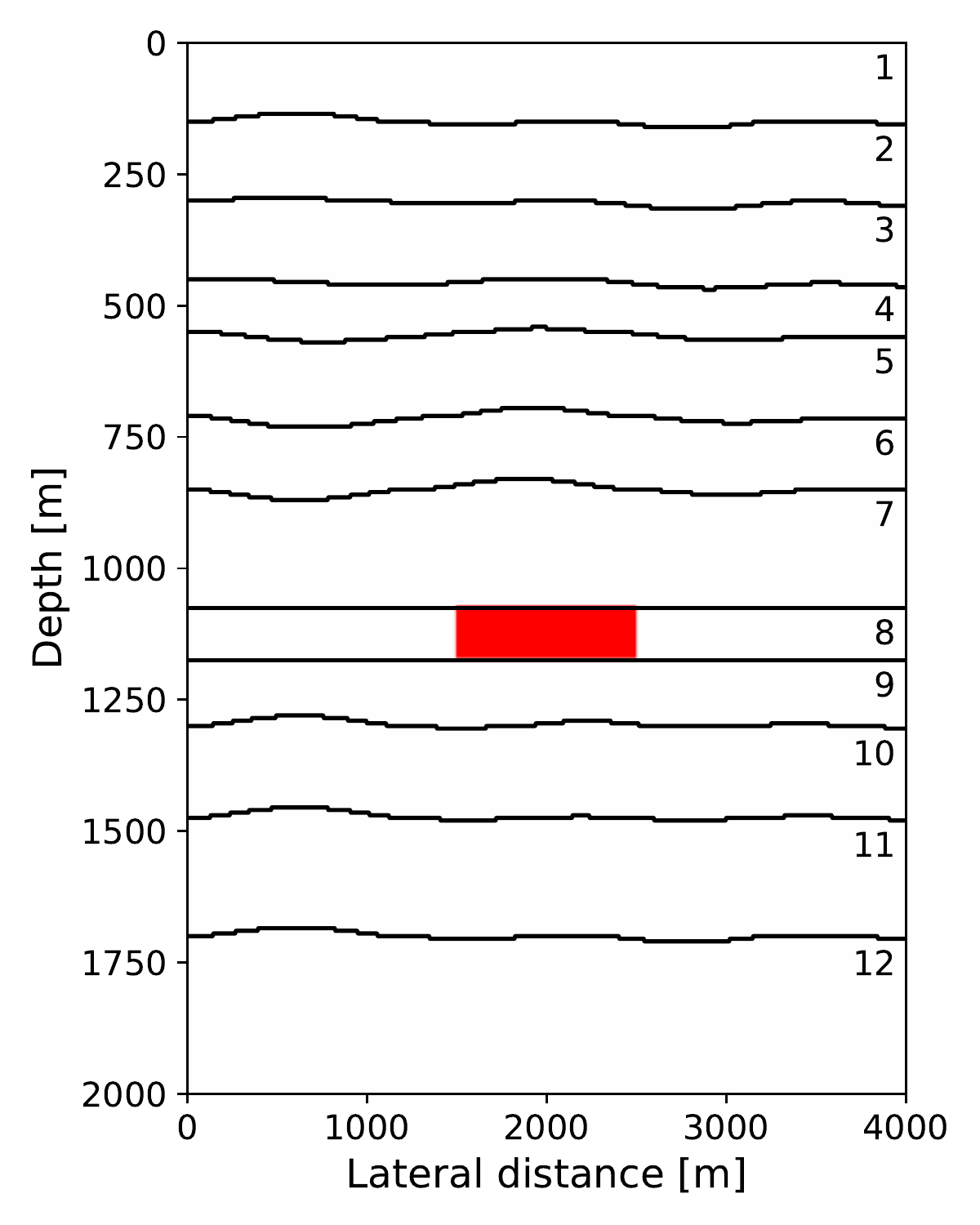}
    \includegraphics[width=.45\textwidth]{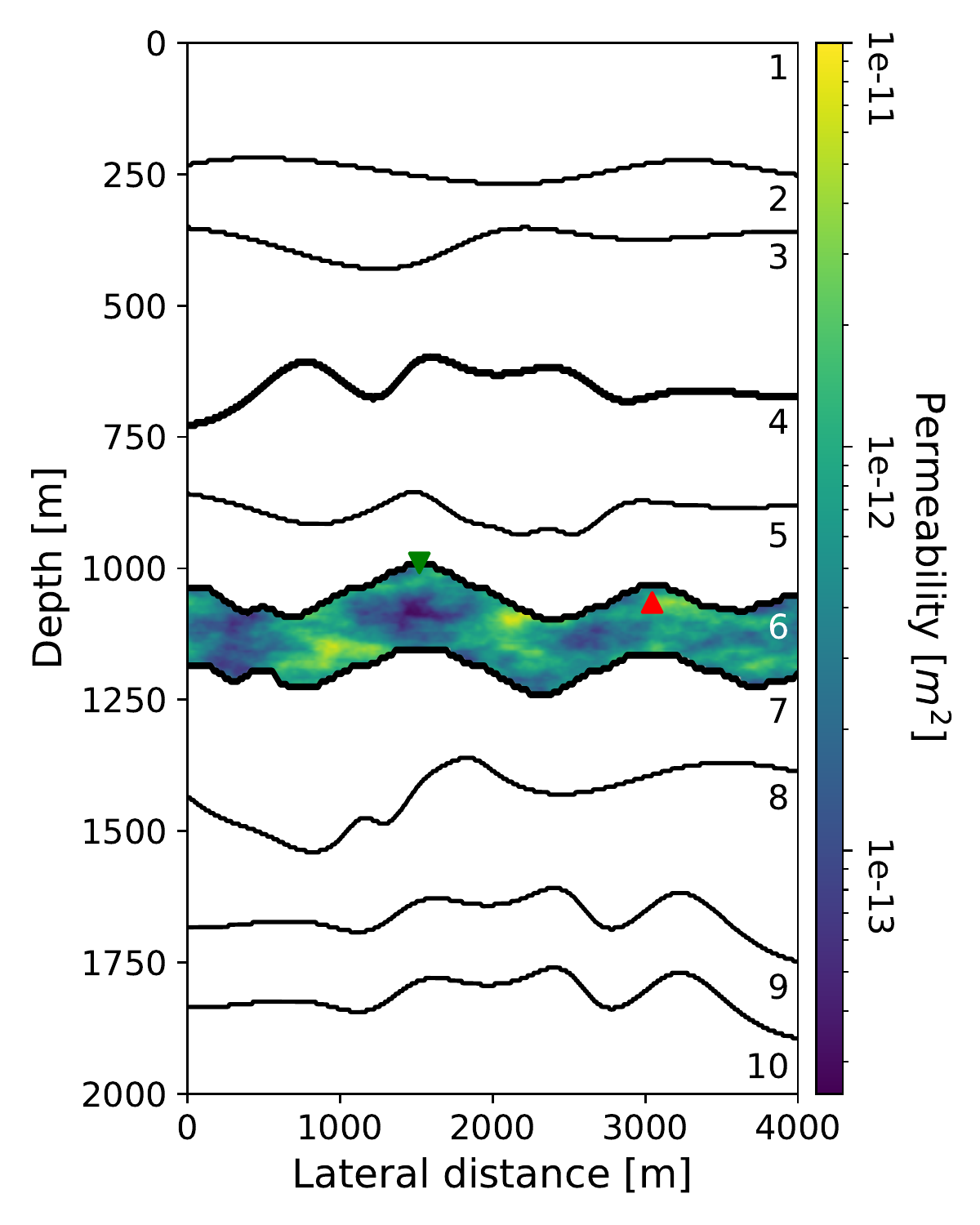}
    \caption{Subsurface model for the first (left) and second (right) numerical example. The reservoir for the first numerical example (in red) has a constant permeability of $1e-13$ m$^2$, and fluid is injected at the left border of the reservoir and produced at the right border. For the second example the reservoir (in colour) has a variable permeability between $2.5e-14$ and $1e-11$ m$^2$. The green and red triangles indicate the location of the injection and production well, respectively (which are now a single point source or sink). Other relevant properties of the numbered layers can be found in \autoref{tab:properties1} for the first model, and \autoref{tab:properties2} for the second model.}
    \label{fig:model1}
\end{figure}

\begin{table}[tb!]
    \centering
    \caption{Properties for each layer and fluid for the first numerical example, layers are displayed on the left of \autoref{fig:model1}. The asterisk indicates the reservoir layer. In this table the elastic parameters are represented by the Poisson ratio $\nu = \lambda/(2\lambda+2G)$ and Young's modulus $E = (G(3\lambda+2G))/(\lambda+G)$. }
    \vspace{.2em}
    \label{tab:properties1}
    \begin{tabular}{|l|r|r|r|r|r|r|r|r|r|r|r|r|}
         \hline
layer \#	&	1	&	2	&	3	&	4	&	5	&	6	&	7	&	8*	&	9	&	10	&	11	&	12	\\
        \hline
$\nu [-]$	&	0.24	&	0.21	&	0.31	&	0.26	&	0.2&	0.3	&	0.25	&	0.15	&	0.19	&	0.25	&	0.3	&	0.2	\\
$E [$GPa$]$	&	10	&	14	&	14	&	15	&	11	&	12	&	6	&	29	&	15	&	32	&	50	&	60	\\
$K_0 [$GPa$]$	&	25	&	20	&	5	&	50	&	50	&	33	&	5	&	30	&	42	&	70	&	10	&	50	\\
$\phi [-]$	&	0.4	&	0.25	&	0.15	&	0.1	&	0.3	&	0.2	&	0.15	&	0.3	&	0.15	&	0.3	&	0.15	&	0.2	\\
$\rho_\text{rock} [$kg/m$^3]$	&	3150	&	2800	&	3500	&	3000	&	2800	&	2800	&	2400	&	3000	&	3100	&	4400	&	5000	&	6000	\\

        \hline
    \end{tabular} 
    \begin{tabular}{|r|r|r|r|r|r|}

        \hline
        fluid & \multicolumn{2}{r|}{$\mu$} & \multicolumn{2}{r|}{$\kappa$} & $\rho_{fluid}$ \\
         & \multicolumn{2}{r|}{[Pa $\cdot$ s]} & \multicolumn{2}{r|}{$[$GPa $^ {-1}]$} & $[$kg/m$^3] $\\
        \hline
        $\alpha$ & \multicolumn{2}{r|}{$1e-3$} & \multicolumn{2}{r|}{$0.5$} & $1035$ \\
        $\beta$ & \multicolumn{2}{r|}{$5e-4$} & \multicolumn{2}{r|}{$1$} & $750 $ \\
        \hline
    \end{tabular} 
\end{table}

\section*{Numerical examples}

A customised fully implicit multiphase poromechanics simulator was designed for this work. The poromechanics part of this simulator was benchmarked on the 1D Terghazi and 2D Mandel problems \citep{Terzaghi1996,Mandel1953}. Furthermore, the two-phase fluid flow simulation was validated using the DARSim Matlab simulator \citep{Cusini2019}. The simulator also includes fluid substitution to find saturated elastic parameters and produce subsurface density and velocity models for seismic modelling. Forward seismic modelling was achieved with an existing finite-difference modeller \citep{Thorbecke2011}, and Marchenko-based isolation of the target response was performed with publicly available algorithms \citep{vanIJsseldijk2023}. In this section two models are considered. The first example is a simple piston-like flow in a homogeneous reservoir with simple overburden. Second, a more heterogeneous model is considered with a highly reflective overburden. 

\begin{figure}[tb!]
    \centering
    \includegraphics[width=\textwidth]{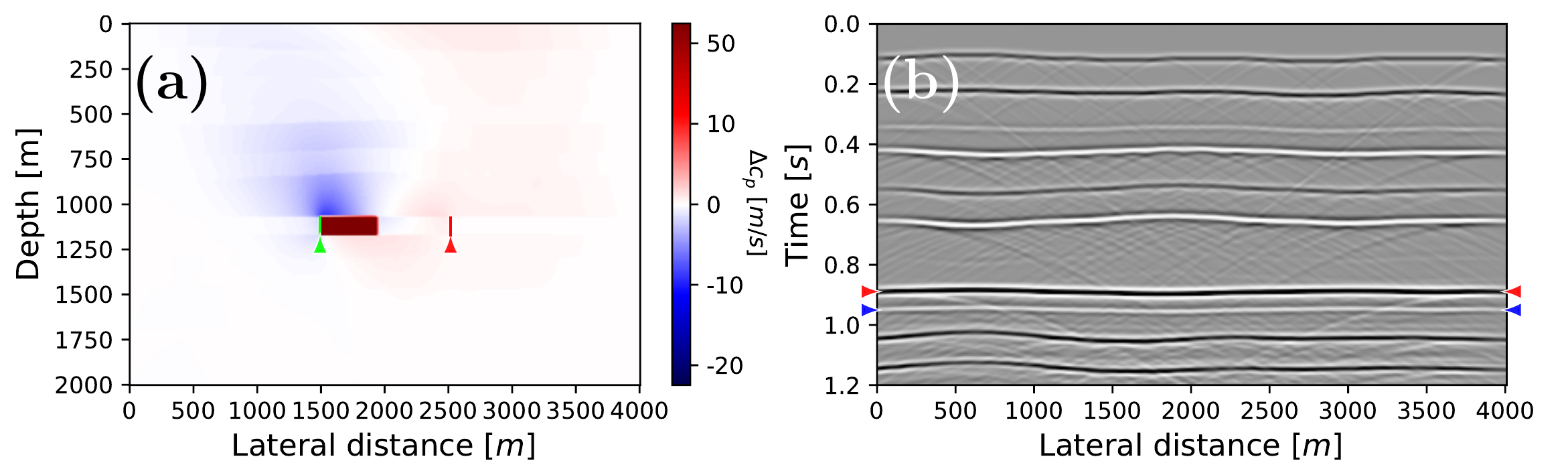}
    \caption{Time-lapse difference in P-wave velocity between 0 and 200 days of simulation (a) and zero-offset reflection response after 200 days of simulating fluid injection (b). In (a) the injection and production wells are marked with the green and red lines, respectively. In (b) the arrows mark primary 1 (red) and primary 2 (blue). These primaries delineate the reservoir and will be used to extract traveltime differences.}
    \label{fig:HomogenDCP}
\end{figure}

\begin{figure}[tb!]
    \centering
    \includegraphics[width=\textwidth]{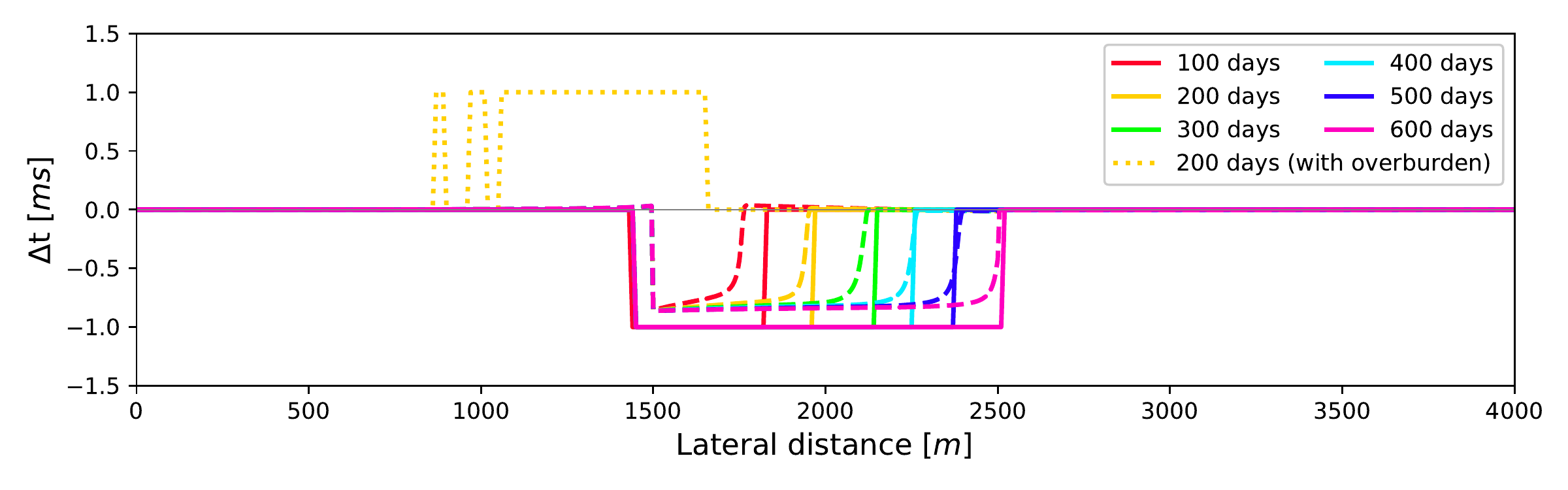}
    \caption{Traveltime differences between the baseline and monitor study at different times of the reservoir simulation for the simple model. The dashed lines represent the actual differences, whereas the solid lines are the differences extracted from the modelled seismic response. To highlight the importance of removing overburden effects, the dotted line shows the traveltime differences after 200 days when primary 2 (below the reservoir) is immediately correlated in the baseline and monitor study. Opposed to this, for the solid lines the overburden effects are first removed using primary 1 (above the reservoir) as reference, and only thereafter the baseline and monitor study are correlated to get the time-lapse changes inside the reservoir.}
    \label{fig:HomogenResult}
\end{figure}

\subsection*{Case 1: Simple model}
The subsurface model for this numerical experiment is shown in \autoref{fig:model1}, and the values for the properties of each layer and fluid can be found in \autoref{tab:properties1}. To start the simulation fluid $\alpha$ is injected on the left side of the reservoir (initially filled with fluid $\beta$) with a constant pressure of $50$ MPa, and the production wells, in line with the right side of the reservoir, have a pressure of $5$ MPa, while the initial pressure in the reservoir is equal to $10$ MPa. The fluid flow is constrained to the reservoir, and roller boundary conditions (i.e. zero normal displacement) are imposed on the four edges of the model. The total simulation time is 600 days, and a seismic survey is modelled at every 100th day. The forward seismic model utilizes a zero-phase wavelet with a flat spectrum between $5$ and $70$ Hz, a time sampling of $4$ ms, and 401 co-located sources and receivers at a spacing of $10$ m. \\
\autoref{fig:HomogenDCP} shows the time-lapse change in P-wave velocity after 200 days compared to the baseline (a), as well as the zero-offset reflectivity modelled at this time step (b). In \autoref{fig:HomogenDCP}a, a decrease in velocity is noted above the injector wells (caused by the time-lapse response of the soft layer on top of the reservoir), whereas the velocity increases above the production wells and inside the reservoir. Inside the reservoir an increase in velocity is noticed due to fluid $\beta$ being replaced with fluid $\alpha$. The changes in P-wave velocity above the reservoir are caused by the pressure change, that is, the increase in pressure due to injection leads to a decrease in velocity; vice versa the pressure decrease above the production wells causes an increase in velocity. Furthermore, primary 1 and primary 2, at the top and base of the reservoir are clearly visible in the seismic reflectivity section, as indicated by the red and blue arrows in \autoref{fig:HomogenDCP}b. Due to this clear visibility, no Marchenko-based isolation is necessary for the simple model. Next, reflectivity is modelled for every 100th day between 0 and 600 days. The initial reflectivity at day 0 is used as baseline study and the subsequent reflection responses are considered monitor studies. \\
After the simulation and forward seismic modelling is finished, the correlation of $P1$ and $P2$ (i.e. the reflection inside the reservoir) is first computed for each of seismic study using \autoref{eqn:timelags}. To further improve the resolution of the results, the correlations are interpolated to $0.5$ ms, by padding the data with zeros in the frequency domain. These correlations are then used to find the traveltime differences between the baseline and monitor studies (\autoref{eqn:timediffs}). The results of this numerical experiment are shown in \autoref{fig:HomogenResult}. The dashed lines in the figure show the reference result based on the time-lapse changes in velocity. The solid lines are time-lapse traveltime differences retrieved with the proposed method. These lines clearly mark the fluid front advancing from left to right in the reservoir. The lines do not perfectly coincide with the reference result due to limitations in the spatial and temporal resolution of the seismic responses, which were measured with a time sampling of $4$ ms and a receiver spacing of $5$ m. Finally, to illustrate the effect of the overburden, the dotted line shows the traveltime differences computed solely from reflection $P2$ of the baseline and monitor data after 200 days. This is in contrast to the solid lines where reflections $P1$ and $P2$ are first correlated, and this correlation is then used to compute the traveltime changes between the baseline and monitor study. This means that the dotted lines do not retrieve the correlation of \autoref{eqn:timelags}, but rather insert reflector $P2$ of the baseline and monitor study directly into \autoref{eqn:timediffs}. All overburden changes are, therefore, included in the dotted line, hence the location of the fluid front can no longer be accurately observed (i.e. it deviates from the reference result). 


\begin{table}[tb!]
    \centering
    \caption{Properties for each layer and fluid for the second numerical example, layers are displayed on the right of \autoref{fig:model1}. The asterisk indicates the reservoir layer.  }
    \vspace{.2em}
    \label{tab:properties2}
    \begin{tabular}{|l|r|r|r|r|r|r|r|r|r|r|r|r|}
         \hline
layer \#	&	1	&	2	&	3	&	4	&	5	&	6*	&	7	&	8	&	9	&	10	\\
        \hline
$\nu [-]$	&	0.23	&	0.24	&	0.16	&	0.25	&	0.27	&	0.25	&	0.4	&	0.31	&	0.24	&	0.35	\\
$E [$GPa$]$	&	6	&	8	&	3	&	11.5	&	2.5	&	7	&	15	&	15	&	20	&	22.5	\\
$K_0 [$GPa$]$	&	25	&	20	&	5	&	50	&	33	&	5	&	30	&	70	&	10	&	200	\\
$\phi [-]$	&	0.4	&	0.25	&	0.15	&	0.25	&	0.1	&	0.3	&	0.3	&	0.15	&	0.1	&	0.15	\\
$\rho_\text{rock} [$kg/m$^3]$	&	2917	&	4333	&	1470	&	4667	&	1333	&	2487	&	4285	&	3176	&	3333	&	4705	\\
\hline
    \end{tabular}
    \begin{tabular}{|r|r|r|r|r|r|}
        \hline
        fluid & \multicolumn{2}{r|}{$\mu$} & \multicolumn{2}{r|}{$\kappa$} & $\rho_{fluid}$ \\
         & \multicolumn{2}{r|}{$[$Pa $\cdot$ s$]$} & \multicolumn{2}{r|}{$[$GPa $^ {-1}]$} & $[$kg/m$^3] $\\
        \hline
        $\alpha$ & \multicolumn{2}{r|}{$1e-3$} & \multicolumn{2}{r|}{$0.5$} & $1035$ \\
        $\beta$ & \multicolumn{2}{r|}{$5e-4$} & \multicolumn{2}{r|}{$1$} & $750 $ \\
        \hline
    \end{tabular}    
\end{table}

\begin{figure}[tb!]
    \centering
    \includegraphics[width=\textwidth]{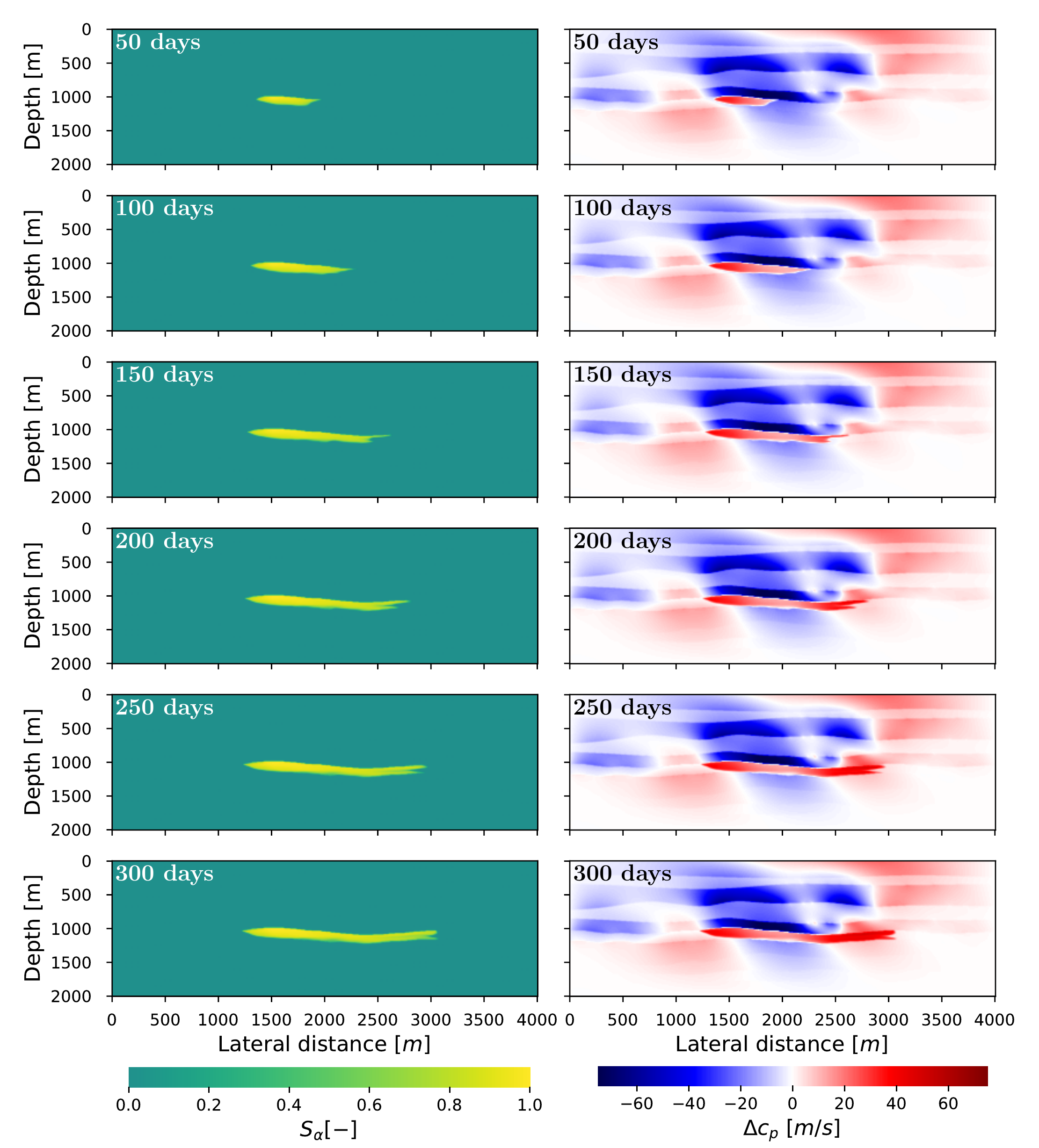}
    \caption{Change in saturation (left column) and P-wave velocity (right column) over time. On the right side the blue and red colours indicate a decrease and increase in velocity, respectively. }
    \label{fig:HeterogenDCP}
\end{figure}

\begin{figure}[tb!]
    \centering
    \includegraphics[width=\textwidth]{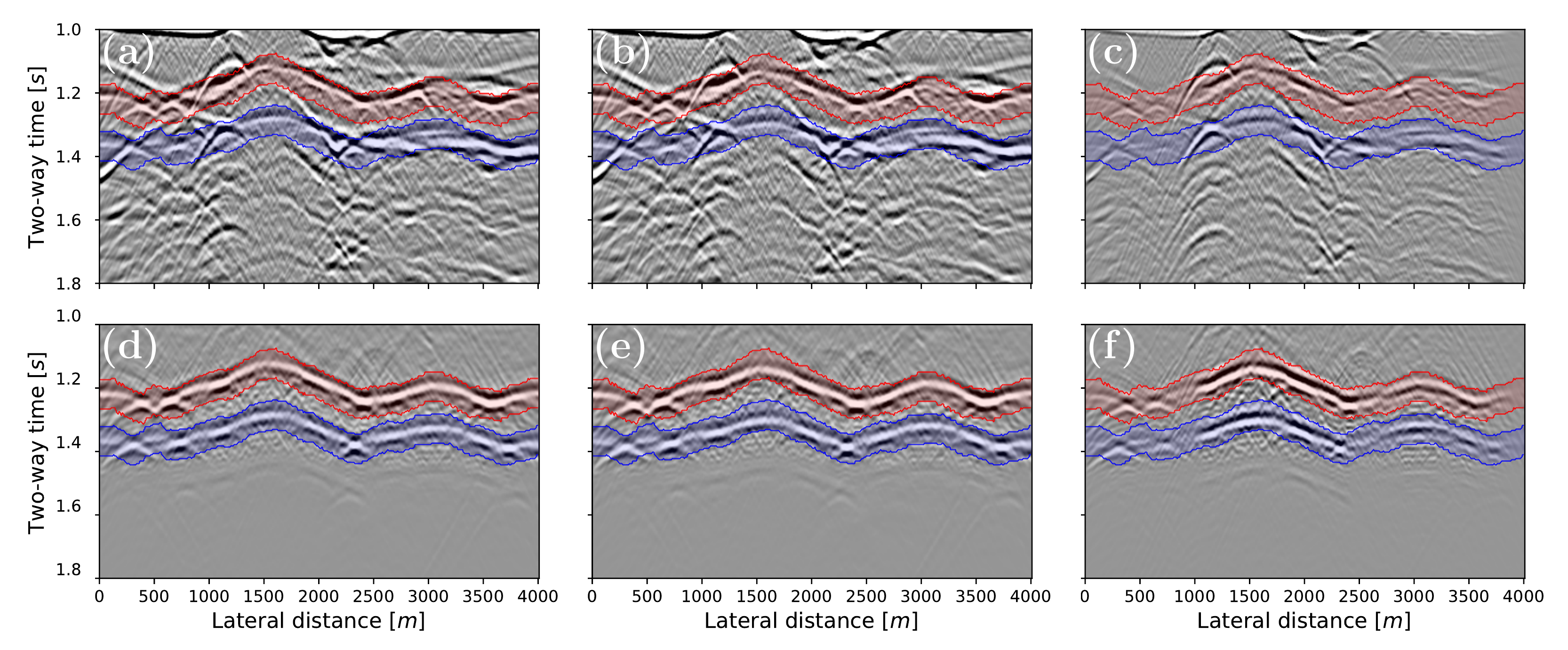}
    \caption{Zero-offset reflectivity at after 0 (left column), 300 days (middle column) and their difference (right column), zoomed in on $1$ to $1.8$ s of recording time. The top and bottom rows show the full ($R_{abc}$) and isolated ($R_b$) response, respectively. The primary at the top of the reservoir ($P1$) is marked in red, and the primary at the base of the reservoir ($P2$) in blue. Note that these primaries are only clearly visible in the isolated responses.}
    \label{fig:zerooffsets}
\end{figure}

\begin{figure}[tb!]
    \centering
    \includegraphics[width=\textwidth]{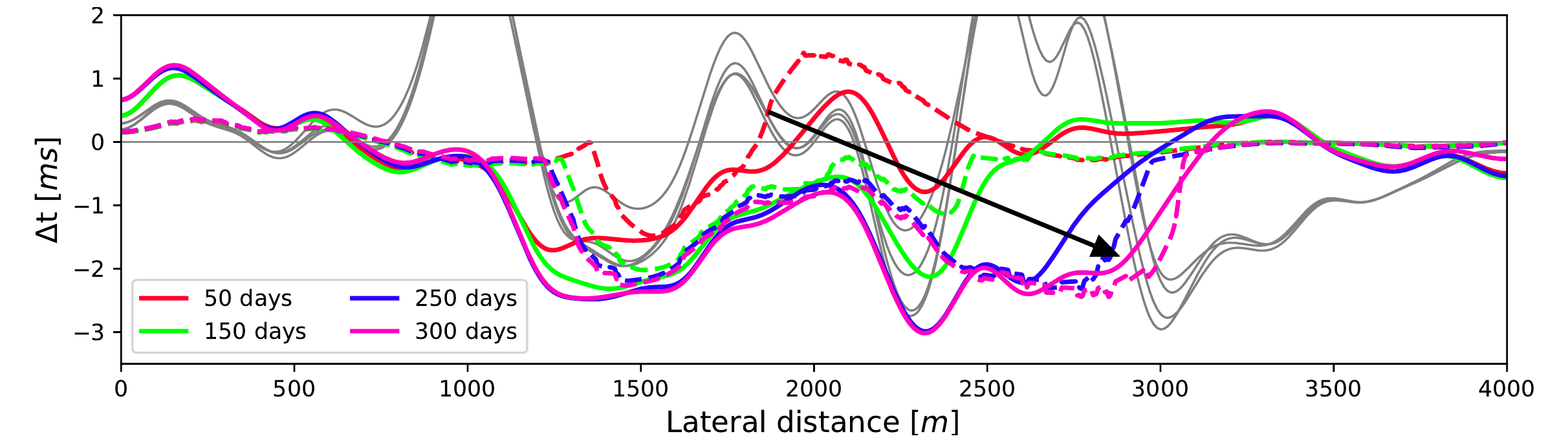}
    \caption{Traveltime differences at different times of the reservoir simulation for the complex model. The dashed lines represent the actual differences, and are included as a reference solutions, whereas the solid lines are the differences extracted from the seismic data after Marchenko-based isolation (i.e. \autoref{fig:zerooffsets}d and \ref{fig:zerooffsets}e). The black arrow indicates the general direction of the fluid front over time. The gray lines show the travel differences extracted from the full response, that is before Marchenko-based isolation (\autoref{fig:zerooffsets}a and \ref{fig:zerooffsets}b).}
    \label{fig:HeterogenResult}
\end{figure}

\subsection*{Case 2: Complex model}
The second numerical experiment examines a more complex model shown on the right in \autoref{fig:model1}, with all layer and fluid parameters listed in \autoref{tab:properties2}. This model contains a highly reflective overburden, designed to produce strong multiple reflections that interfere with the reservoir response \citep{vanIJsseldijk2021}. Additionally, this also means that the overburden yields a strong response due to geomechanical changes. Furthermore, the reservoir is no longer rectangular, instead it has a wave-like structure with variable permeabilities between $2.5e-14$ and $1e-11$ m$^2$. This permeability is pseudo random, generated using Perlin noise that allows for a somewhat coherent distribution \citep{Perlin1985}. As shown in \autoref{fig:model1}, a single injector well ($50$ MPa) is located at the top of the reservoir at $1500$ m lateral distance, alongside with a production well at $3000$ m with a pressure of $5$ MPa. The boundary conditions are the same as in the simple model. Every 50th day of simulation the reflection response is modelled, the total simulation time is 300 days. The seismic modelling uses the same parameters as in the first numerical example, except for the spectrum of the source wavelet, which is flat between $5$ and $50$ Hz. \\
The evolution of the saturation is shown in the left column of \autoref{fig:HeterogenDCP}. This figure also displays the changes in velocity for 6 monitor time steps, which shows that the time-lapse changes outside the reservoir overpower the changes inside the reservoir. Again, the velocity decreases and increases above the injection and productions wells, respectively. \autoref{fig:zerooffsets} displays the zero-offset seismic sections of the initial baseline and final monitor studies at 0 and 300 days, respectively. \autoref{fig:zerooffsets}a and \ref{fig:zerooffsets}b show that the two primary reflections (marked in red and blue) are obscured by overburden events. As seen in \autoref{fig:zerooffsets}c the time-lapse differences from the reservoir are masked by overburden effects. Consequently, it is beneficial to isolate the reservoir response, contrary to the first example. The results of this isolation are shown in \ref{fig:zerooffsets}d, \ref{fig:zerooffsets}e and \ref{fig:zerooffsets}f. The desired events (i.e. $P1$ and $P2$) are now revealed in the reflection response. Similarly, the isolation is applied to the remaining monitor surveys as well as the baseline survey. Next, the correlations are interpolated from $4$ to $1$ ms via the frequency domain for improved resolution. Subsequently, the correlations between $P1$ and $P2$ are computed, which will be used to find the time differences of the reservoir layer. \\
Once again, the time-lapse traveltime differences inside the reservoir are computed according to \autoref{eqn:timediffs}; the results of this procedure are plotted with solid lines in \autoref{fig:HeterogenResult}. As before, the dashed lines depict the reference result calculated based on the velocity changes in \autoref{fig:HeterogenDCP}. However, their behaviour is a lot more complex than the previous results. The solid, coloured lines represent the traveltimes difference computed from the correlations of $P1$ and $P2$ of the baseline and monitor surveys. Lastly, the gray lines show the results of computing the time-differences from the full medium responses (i.e. without isolation, \autoref{fig:zerooffsets}a and \ref{fig:zerooffsets}b), these results clearly deviate from the reference solution due to overburden effects appearing in the selection window of the primaries. Even though the computed differences somewhat agree with the reference solution, the match is significantly poorer than for the simple model. Especially around $2200$ m the results strongly differ from the reference solution, as all 4 solid lines underestimate the reference indicated with the dashed lines. This underestimation is also observed in the gray lines in the background, and could, therefore, indicate that some remainder of overburden effects is still present in the isolated response. Nevertheless, the results after applying the reservoir isolation (i.e. the solid coloured lines) show a clear improvement in recovering the trend of the fluid movement compared to the solid gray lines. Further improvements may be possible using a migration technique to collapse the diffractions to their true location, thus improving the spatial resolution of the data. 

\section*{Discussion}

In the previous section it was shown that poromechanical modelling can add valuable insights to seismic reservoir monitoring, specifically because overburden changes are predicted together with fluid flow. In this section possible improvements upon both the poromechanical and the seismic part of the method are discussed. \\
Firstly, poromechanical simulations are computationally expensive, which sets practical limitations on their use. Currently, the simulation takes around 8 hours, when using 8 CPUs on the Delft High Performance Computer \citep{DHPC2022}. However, this time dramatically increases on a grid with finer discretization. One solution to this problem is to apply a preconditioner to improve convergence of the equations, allowing to still achieve high resolution simulations \citep{white2019}. Alternatively, a multiscale approach could be used in order to limit the size of the problem, thus speeding up the simulations \citep{Cusini2016,Hosseinimehr2018,Sokolova2019}. \\
Additionally, the simulator could be improved by extending the formulations to include fractures. This is especially relevant as fractures can unexpectedly block the fluid or bypass impermeable zones. Another feature that is currently missing from the simulator is the ability to model cyclic storage, which is required to accurately monitor temporary storage of hydrogen or other gasses in the subsurface \citep{Kumar2021}. It would also be interesting to include a third fluid to the simulator, in order to account for solution and dissolution of gasses into fluid due to the pressure changes inside the reservoir. \\
In the current seismic study, only timelapse time-differences were considered. Future developments should also consider seismic amplitude variations of the signal. Recent work investigated how angle-dependent amplitude information can be retrieved with the Marchenko method \citep{Alfaraj2020}. Ideally, a combination of both amplitude and traveltime differences is used to recover the dynamic fluid parameters from the seismic data \citep{Trani2011}. \\

\vspace*{-.05cm}
\section*{Conclusion}
This work developed a multiphase poromechanical simulator that is tightly coupled with a seismic survey simulator, through the update of the seismic parameters. This integrated simulator allows to instantly resolve time-lapse changes not only inside the reservoir but also in its overburden. The simulator was used to test the feasibility of a novel methodology to extract time-lapse travel time changes after Marchenko-based isolation of the reservoir response. This methodology solves the repeatability issue of time-lapse seismic surveys by identifying and utilizing a reference reflection above the reservoir. Future developments should focus on inverting the seismic time-lapse changes back to the dynamic reservoir properties, in order to close the loop between reservoir simulations and seismic monitoring. \\
These results are a significant step to achieve higher resolution monitoring of subsurface reservoirs. A better understanding of fluid flow in these reservoirs will improve predictions of underground processes related to geothermal energy, subsurface storage of gasses like hydrogen, and sequestration of CO$_2$.

\section*{Data availability}
The poromechanical simulator is publicly available here: \href{https://github.com/Ohnoj/PoroMechanics-Simulator}{https://github.com/Ohnoj/PoroMechanics-Simulator}. Algorithms associated with the seismic forward modelling and Marchenko-based isolation can be accessed via the following URL: \href{https://gitlab.com/geophysicsdelft/OpenSource}{https://gitlab.com/geophysicsdelft/OpenSource}.

\bibliography{Thesis}

\section*{Acknowledgements}
The authors are grateful for insightful comments and discussion with the members of ADMIRE research group and user committee. The research of J.v.IJ. and K.W. was funded by the European Research Council (ERC) under the European Union’s Horizon 2020 research and innovation program (grant agreement no. 742703). H.H. was sponsored by the Dutch National Science Foundation (NWO) Talent Programme ViDi Project “ADMIRE” (Grant number 17509).

\section*{Author contributions statement}

J.v.IJ.: poromechanics and seismic methodology, software, writing—original draft. H.H.: poromechanics methodology, supervision, writing—review and editing, grant acquiring, K.W.: seismic methodology, supervision, writing—review and editing, grant acquiring.

\section*{Additional Information}
The authors declare no competing interests.

\newpage


\end{document}